# Cause of Chirality Consensus

Salla Jaakkola[a], Vivek Sharma[b], Arto Annila[a,b,c*]

[a] *Department of Biological and Environmental Sciences, University of Helsinki, Viikinkaari 3, FI-00014, Finland*
[b] *Institute of Biotechnology, University of Helsinki, Viikinkaari 1, FI-00014, Finland*
[c] *Department of Physical Sciences, Gustav Hällströminkatu 2, University of Helsinki, FI-00014, Finland*

**Abstract**: Biological macromolecules, proteins and nucleic acids are composed exclusively of chirally pure monomers. The chirality consensus appears vital for life and it has even been considered as a prerequisite of life. However the primary cause for the ubiquitous handedness has remained obscure. We propose that the chirality consensus is a kinetic consequence that follows from the principle of increasing entropy, i.e. the 2$^{nd}$ law of thermodynamics. Entropy increases when an open system evolves by decreasing gradients in free energy with more and more efficient mechanisms of energy transduction. The rate of entropy increase is the universal fitness criterion of natural selection that favors diverse functional molecules and drives the system to the chirality consensus to attain and maintain high-entropy non-equilibrium states.

## 1. INTRODUCTION

In 1848 young Louis Pasteur observed optical activity in a natural extract of tartaric acid solution whereas a racemic solution, with both enantiomers present, was devoid of it. The study revealed also a preferred handedness as the two enantiomers formed separate crystals. This well-known discovery exposed homochirality that is characteristic of all living systems. Biological macromolecules most notably nucleic acids and proteins that are involved in numerous activities are chirally pure.

The basis for the broken symmetry of life has been sought from intrinsic asymmetries in fundamental physical processes [1]. The very thought was already favored by Pasteur but inherent preferences for one or the other enantiomer due to parity violating energy differences between enantiomers [2,3] have been estimated to be minute, on the order of $10^{-13}$ Jmol$^{-1}$ [4,5]. Not disregarding such processes, their enhancement seems necessary when an emerging handedness is constantly jeopardized by spontaneous racemization [6]. Amplification scenarios have been suggested, including extra terrestrial phenomena [7,8], to yield a chirally pure prebiotic medium for life to emerge. Thus the handedness has been associated with the origin of life being a singular event [9] where for example a random choice or enantioselectivity in adsorption of amino acids on mineral surfaces dictated the outcome for the entire biota [10,11] Alternatively it has been proposed that initially both enantiomers formed, coexisted and competed with each other finally ending to the unanimous victory by one of them but it has remained unclear how natural selection operates on molecules without mechanisms of replication. Nevertheless, autocatalytic reactions may spontaneously resolve to a pure-hand state [12,13], and residues may continue to oligomerize into homopolymers [14]. Symmetry breaking processes of non-equilibrium systems [6] have also been observed in crystallization [15,16]. On the basis of these seemingly exceptional cases the significance of kinetics as the ultimate cause of chirality has remained uncertain because the emergence of handedness has not been associated with a fundamental physical law. The cause of chirality consensus has been the subject of numerous hypotheses but to our knowledge an explanation that would relate also to the origin and characteristics of life itself has not been found.

In this study we find the ubiquitous handedness to follow from the principle of increasing entropy using the recently formulated statistics of open systems [17]. The formulation essentially expresses the theory of evolution by natural selection [18] in terms of physical chemistry. The theory has been applied to explain e.g. why many populations of animals and plants as well as gene lengths display skewed nearly log-normal distributions and sigmoid cumulative curves [19]. Here we show that under an influx of external energy the 2$^{nd}$ law of thermodynamics compels evolution of an increasingly larger chemical system toward a ubiquitous chirality consensus. According to thermodynamics the homochirality standard, the ubiquitous signature of life, emerged to allow rapid synthesis of diverse functional structures, catalysts that are mechanisms of energy transduction to attain and maintain high-entropy non-equilibrium states. We do not question kinetic scenarios and particular mechanisms that have been presented earlier to describe *how* homochirality might have risen but point out that kinetics takes the direction of increasing entropy to explain *why* handedness emerged during evolution.

## 2. THE DRIVE TO HIGH-ENTROPY NON-EQUILIBRIUM STATES

The principle of increasing entropy accounts for many spontaneous phenomena in nature e.g. heat flow from hot to cold and for molecular diffusion down along concentration gradients. Also chemical systems evolve via reactions toward the most probable state, i.e. the chemical equilibrium, by flows of matter and energy between reactants and products [20,21]. All these natural processes evolve toward high-entropy states by consuming thermodynamic forces i.e. potential energy differences using various mechanisms to duct energy. Entropy $S$, derived from statistical physics, is a mere logarithmic probability (ln$P$) measure to compare attainable states of a system. All systems are on their way toward more probable states. The probable course directs down along the potential energy difference between the evolving system and its surroundings. An exoergic reaction is probable when energy can be dissipated to surroundings. Likewise an endoergic reaction is probable when it couples to external energy. Both exo- and endoergic reactions are manifestations of 2$^{nd}$ law that levels differences among energy densities.

It takes interaction mechanisms, e.g. functional structures, to channel energy to the system or out of the system in order to diminish free energy gradients. Structure associated local losses in degrees of freedom are well compensated by functionalities that allow the system as a whole to move toward more probable states by leveling potential energy differences. High-entropy non-equilibrium state of a living system is attained and maintained by replenishing diverse metastable mechanisms that channel energy to the system from its exterior. Thus entropy *increases* also in the processes of life [17], in contrast to common but erroneous conceptions. As given below entropy is a measure of energy dispersal [22] as being a probable course, not a plain merit of disorder [23,24].

Entropy $S$ of an open system undergoing chemical reactions as function of time $t$ is given by (see reference 17 for the derivation)

$$S = R\sum_{j=1} \ln P_j \approx \frac{1}{T}\sum_{j=1} N_j \left(\sum_k \mu_k + \Delta Q_{jk} - \mu_j + RT\right) \quad (1)$$
$$= R\sum_{j=1} N_j \left(\frac{A_j}{RT} + 1\right)$$

where $\mu_k/RT=\ln[N_k\exp(G_k/RT)]$ and $\mu_j/RT=\ln[N_j\exp(G_j/RT)]$ denote chemical potentials of substrates $N_k$ and products $N_j$ relative to the average energy $RT$ per mole. Evolution takes the direction toward increasingly more probable states ($dS/dt > 0$) using the potential energy differences that are in the case of chemical reactions referred to as affinities $A_j = \Sigma\mu_k+\Delta Q_{jk}-\mu_j$ [20]. An external quantum of energy $\Delta Q_{jk}$ that couples to a reaction contributes to the affinity as a substrate. Diverse compounds that emerge from various reactions are index by $j$. They are repositories $\mu_j$ of chemical energy but may also act as energy transduction mechanisms to facilitate growth of entropy i.e. evolution. Thus all interacting compounds contribute to the rate of entropy increase [17]

$$\frac{dS}{dt} = \sum_{j=1} \frac{dS}{dN_j}\frac{dN_j}{dt} = \frac{1}{T}\sum_{j=1} \frac{dN_j}{dt}\left(\sum_k \mu_k - \mu_j + \Delta Q_{jk}\right) \quad (2)$$
$$= \frac{1}{T}\sum_{j=1} v_j A_j.$$

that manifests itself as diverse flows $v_j$ of matter via chemical reactions.

The flow rates are customarily modeled by the mass action law [20,22,25] but are according to Eq. 2 proportional to thermodynamic force $A_j/RT$ [17]

$$v_j = \frac{dN_j}{dt} = r_j \frac{A_j}{RT} \qquad (3)$$

to satisfy the balance equation $v_j = -\sum v_k$. In other words the flow to the product pool equals the flow from the substrate pools when taking into account also the acquired or dissipated quanta. The rate constant $r_j$ depends on mechanisms e.g. catalysts. Increasingly more efficient mechanisms provide access to states that are increasingly higher in entropy where the difference in energy-densities between the system and its exterior is smaller.

The open system continues to evolve via reactions further when new mechanisms emerge to acquire more matter and energy to the natural process. When no new resources appear, the most probable state will be ultimately reached. Then the affinities vanish and the chemical potentials across all reactions are equal [17]

$$dS/dt = 0 \Leftrightarrow N_j = \prod_k N_k \left[ \exp\left( \Delta Q_{jk} - \Delta G_{jk} \right) / RT \right] \qquad (4)$$

to reveal that the energy influx $\Delta Q_{jk}$ supports the non-equilibrium high-entropy states. The non-equilibrium concentrations of metastable ($j > 1$) compounds $N_j$ are above those of the equilibrium determined solely by $\Delta G_{jk}$. This dependence of the most probable state on external conditions is consistent with the LeChâtelier's principle.

Compounds with functional, e.g. catalytic, properties are very important for the system to reach and maintain high-entropy non-equilibrium partitions. In nature many powerful mechanisms associate with functional structures as is expressed by the catchphrase 'structure-activity relationship'. When diverse mechanisms draw from the same limited resources, the natural selection takes place on the basis of the rates entropy increase. In other words flows of matter direct via efficient mechanisms. According to eqs. 2 and 3 increasingly more functional structures boost kinetics of energy transduction thus resulting pathways for rapid entropy increase.

### 3. THE ORIGIN OF UBIQUITOUS HANDEDNESS

The principle of increasing entropy by decreasing free energy gradients, given by eqs. 1 – 4, allows us to reveal a potential cause of chirality consensus. For simplicity we consider an initial state, presenting prebiotic conditions, consisting only of achiral primary compounds $N_1$ where the index 1 denotes various stable (abiotic) base compounds, e.g. water and carbon dioxide. However, we do not make any hypotheses what these base compounds might have been.

If the base compounds couple to any source of external energy, e.g. to sunlight, there is the driving force of evolution. In the quest to level the potential energy difference a racemic system will arise when the primary compounds react by absorbing external quanta and yield chiral secondary compounds $N_2^L$ and $N_2^R$ where the indexes $L$ and $R$ are symbolic without any reference to actual optical properties. We will consider neither the mechanisms of syntheses nor speculate about nature of secondary compounds. We will only elucidate the probable course of evolution when diverse compounds emerge. Notably, according to eq. 1 the compounds are distinguished as chiral in their mutual interactions, not by a fictitious external observer who may even identify enantiomers that are not differentiated in the system's interactions. When some chiral compounds have appeared, e.g. only via random syntheses, eqs. 1 – 4 allow us to deduce and exemplify the evolution to the homochiral outcome, see Fig. (**1**).

When diverse compounds emerge, some of them may possess some emergent catalytic properties. Some homochiral polymers may act as catalysts of the particular polymerization reaction that ligates homochiral constituents whereas a mixed-hand compound may not be as an effective catalyst for *all* reactions that consume base constituents of varying handedness. Consequently, the homochiral catalysts are favored in attaining high-entropy states as being fewer and faster over those many heterogeneous catalysts that are required in syntheses of diverse mixed-hand compounds. Due to the virtuous circle for increased entropy rates, matter and energy will flow at increasing rates in the two systems corresponding to the two chirally pure branches, whereas systems with mixed-hand compounds will lag more and more behind, see Fig. (**1**).

This overall *kinetic* scenario has been reasoned also earlier [26,13], however it is first here founded on the fundamental principle of increasing entropy (Eq. 2). The entropy increase rate (Eq. 2) leading to the chiral convention is the self-reinforcing character of natural selection. The imperative is independent of mechanisms and generally applicable to all kinetics (Eq. 3).

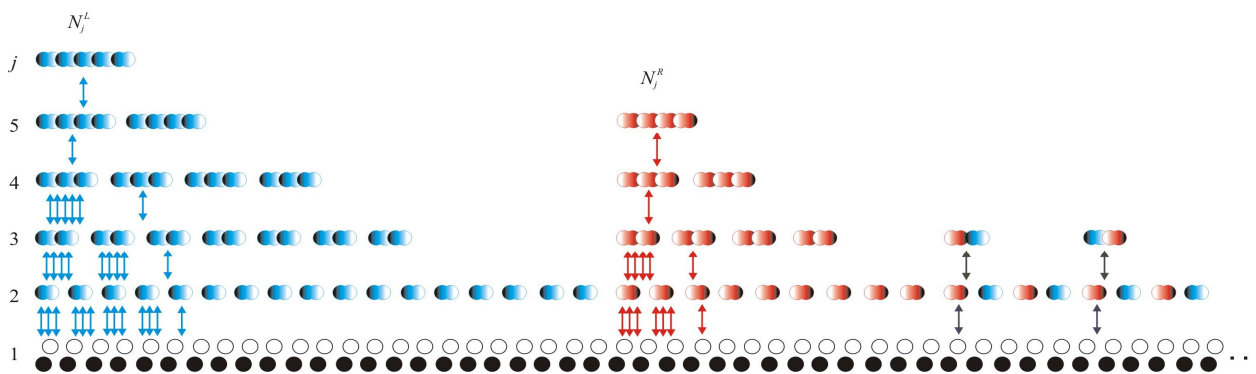

**Fig. (1).** Energy level diagram of molecular diversity among compounds of pure and mixed handedness. An influx of external energy provides high energy potential that drives matter to organize via chemical reactions from stable achiral basic constituents (●○) at the base level ($j = 1$) in diverse metastable chiral entities ($N_{j>1}$) at higher levels ($j > 1$) either by spontaneous transitions (↕) or by reactions that are catalyzed by previously formed compounds (↕↕↕). Blue color marks the left-handed and red the right-handed enantiomers. The two homochiral systems are shown in plain colors of blue and red whereas the mixed-hand systems are mixed with colors of blue and red. The various systems will continuously compete for matter and energy to increase their entropy when drawing from the common pool of base constituents that keep replenishing as the metastable compounds break apart. Those systems where high-$j$ entities happen to catalyze syntheses of their own low-$j$ precursors will reach furthest by evolving toward chirally pure states. In contrast those systems with mixed-hand compounds will not contain all conceivable catalysts to produce all precursors. Therefore the homochiral systems shown to occupy high levels will grow faster and larger than the mixed systems. However, the coexistence of two chirally pure systems is not stable. Fluctuations render one of the systems slightly larger and thus slightly faster in consuming free energy gradients. Inevitable it will supersede its counterpart by including more and more matter and energy in its progression toward higher entropy. Therefore the blue system is shown larger than the red system.

## 4. EMERGING CHIRAL CONSENSUS

The rate of synthesis is a very important criterion to reach and maintain the non-equilibrium high-entropy states. When energy transduction mechanisms fail e.g. due to spontaneous degradation, the coupling to the external source will break and the leveling of potential energy differences will cease. Increasingly larger energy transduction machinery will commit to one kind of handedness, i.e. to a standard to acquire and maintain more efficient mechanisms to decrease potential energy differences.

Evolution to chirally pure states has been simulated earlier using kinetics of the mass action law [20]. However, these differential equations are near-equilibrium approximations for a set of equations that identify the free energy differences including external energy as the thermodynamic driving forces also for non-equilibrium states. According to Eq. 4 the equilibrium and the stationary non-equilibrium state differ only by the external energy $\Delta Q_{jk}$. In the presence of external energy the chemical system emerges with mechanisms that facilitate evolution by diminishing potential energy differences.

We simulated an emerging chiral consensus by catalyzed polymer endoergic syntheses $N_{j-1} + N_1 \rightleftarrows N_j$ that coupled to external energy ($\Delta Q$) and exoergic degradations $N_j \rightleftarrows jN_j$. The syntheses were assigned higher rates $r_j$ progressively with $j$, i.e. longer polymers were modeled as increasingly better catalysts. In a self-similar manner the reaction mechanism itself was considered as a result of a natural process by assigning catalytic rates $r_j \propto \Sigma\mu_n\exp(-\mu_n)$, $n \leq j$ i.e. weighting chemical potential $\mu_n$ by its thermodynamic partition. The degradation rates of metastable products were modeled non-catalytic and proportional to the chemical potential differences without the external energy. The functional forms of $r_j$, the total amount of base constituents and the flow of external energy govern the time course of evolution. We emphasize that specific forms of syntheses and degradations are not important because all systems evolve to use mechanisms of energy transduction that provide high rates of entropy increase. As flows of matter channel via increasingly more efficient mechanisms those early rudimentary and less effective reaction mechanisms will inevitably go extinct.

The evolution was programmed simply as a for-loop of time steps for syntheses and degradations according to the equation 3. The two sets of reactions corresponding to the two chiral systems of homopolymers were otherwise identical but their rates $r_j$ were varied randomly (up to 10%) during each time step. Chemical potentials were recalculated after every step of synthesis and degradation to obtain updated free energy differences. At each time step the entropy according to Eq. 1 and the degree of handedness were calculated from the concentrations of compounds. Entropy and the degree of handedness were found to increase during the evolution, see Fig. (**2**), but they were not used in any way to direct the courses of reactions.

We find that systems with higher catalytic rates, higher energy intake, will grow larger and reach higher degrees of chirality consensus and higher values of entropy. In our simulations the system evolves to one or the other form of handedness as a result of early random fluctuations in syntheses. During the early part of the simulation the degree of chirality is low and fluctuates. Small systems do not even resolve to a state of ubiquitous chirality whereas increasingly larger systems become during evolution increasingly more committed to the standard. However there is no identifiable single event that would dictate the choice of chirality, the ubiquitous handedness emerges as a consequence of natural process.

Often autocatalysis is associated with the kinetic scenarios that emerge with a chirality consensus. However according to the principle of increasing entropy all mechanisms of energy transduction will move the system toward more probable states. The principle holds also for hypercyles [27] and catalytic closures [28] that are according to the Lyapunov criterion more stable than mere autocatalytic systems. The chirality consensus follows from the quest to diminish free energy gradients in the most probable way as expressed by eqs. 2 and 3 however the specific mechanisms are not exposed by the principle. If a way to increase entropy emerges it will be taken. Later if another way appears but giving a higher $dS/dt$ rate than the former route, the latter will dominate and eventually take over all flows of matter and energy. Summations in eqs. 1 and 2 are deliberately without upper bounds to signify that new mechanisms of energy transduction may appear in an open system and redirect its course of evolution. A large system with numerous base constituents may come up with increasingly more efficient mechanisms over the eons.

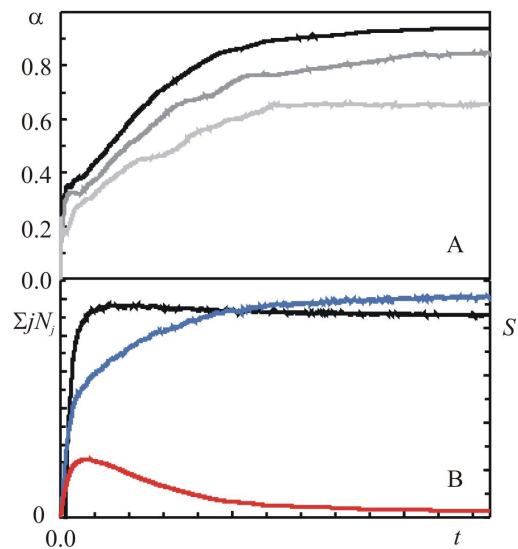

**Fig. (2).** (A) Degree of chirality $\alpha$ as a function of time $t$ obtained from simulation of evolution for three example systems that differed by their relative sizes (4, 2, 1 on the scale from black to grey). According to eq. 3 polymerization reactions $N_1+N_{j-1} \rightleftarrows N_j$ were simulated. The Gibbs free energy levels $G_j$ for the compounds $j$ were modeled as equidistant increasing with $j$ from $G_1 = 0$. The chemical potentials were obtained from $\mu_j = RT\ln[N_j\exp(G_j/RT)]$ assuming a constant temperature $T$. Catalytic rates were assigned progressively with $j$ to chiral compounds as described. The degree of chirality $\alpha = (N_j^L - N_j^R)/(N_j^L + N_j^R)$ contains all chiral units $\Sigma jN_j$ as free or polymerized. Evolution of increasingly larger systems reached states with increasingly higher degree of handedness. (B) Random fluctuations during early syntheses perturbed the delicate even situation for various chiral compounds (blue and red) and dictated the choice of chirality when more and more powerful catalysts of energy transduction appeared in the system shown here for the largest simulated system. Entropy $S$ (black) increased when various chiral compounds (blue and red) emerged from syntheses.

Now that the formula of entropy is known also the Lyapunov function is known and the powerful mathematical methods of non-linear dynamics [29] are at our disposal. It can be shown by the Lyapunov stability criterion that systems with opposite handedness cannot coexist for a long time when they draw from the common pool of base constituents, see Fig. (**2**). The second variation of entropy and its time derivate with respect to fluctuations $\delta N_1$, i.e. $S(\delta N_1) < 0$ and $dS(\delta N_1)/dt > 0$, reveal that a such situation is not stable and co-existence times have been calculated [30]. This rule, known also by the competitive exclusion principle [31], can be understood by the index theory as well [29]. Fluctuations will make one or the other chiral branch slightly larger and thus also catalytically more potent to channel increasingly larger flows of energy from high-energy sources. As a result the base constituents will begin to flow into the faster and hence to the increasingly larger system from the smaller and less potent system. The entropy advantage of a homochiral system increases with the size because more and more powerful catalysts appear in a larger and larger system, see Fig. (**2**). Thus we reason that the primordial world turned to the ubiquitous handedness in a logistic manner as more and more matter and energy began to flow via the homochiral mechanisms to propel the growth of entropy.

In mathematical terms of non-linear dynamics gradient systems ($A_j \neq 0$) evolve by bifurcations that signal for emergence of new energy transduction mechanisms. These more powerful mechanisms consume free energy gradients and associate with fixed points, e.g. molecular species that occupy basins in the free energy landscape. Owing to the rate of entropy increase criterion for fitness some other species with less powerful mechanisms may even go extinct. Eventually all systems will reach the steady state ($dS/dt = 0$) when no new fixed points will appear and no old ones will disappear to deform the entropy landscape further and to cause net surges between the points. Then the total index of a closed curve about the system equals +1. Evolution has come to its end and remaining motions, e.g. fluctuations or oscillations, are closed orbits [29].

## 5. DISCUSSION

It has been shown theoretically as well as experimentally that autocatalysis or catalysis may lead to the homochirality [11,12,13,32]. Thermodynamics of open systems is consistent with this reasoning and observations, however, we like to point out that direct autocatalysis in self-replication is not a stringent requirement of handedness but the rate of entropy increase, as the universal fitness criterion, drives non-equilibrium systems to pure-hand catalytic hypercycles and closures. Also a second-order-like phase transition to homochirality has been shown to depend on concentrations of reactants and reaction rates [20] in accordance with eqs. 1–3 as well as with the simulations shown in Fig. (**2**).

The autocatalytic reactions that emerge with chirally pure states are often regarded as mere exceptions but according the statistics of open systems, most simple systems do not have enough ingredients and external energy to emerge with structures that would be noticeably functional in recruiting more matter and energy to the natural process. Then evolution stalls at its first steps.

Furthermore experimental setups are usually closed and prevent the system from coupling to external supplies hence also from much of evolution. Likewise when a simple system is initially prepared to a chirally biased-state, it will soon resolve to an equilibrium racemic solution by dissipation because it is without much of energy transduction mechanisms to recruit external energy to maintain the chirally pure non-equilibrium state. First a large open system may have enough matter and energy to evolve toward highly functional structures with ubiquitous standards such as homochirality to increase entropy. Apparently the Miller-Urey experiment [33] and those processes that produced chiral compounds found in meteorites [34] did not involve enough matter, energy or time for the natural process to develop ubiquitous conventions, signatures of rapidly and extensively diversifying matter to support rapid entropy increase.

We find from the principle of increasing entropy by decreasing free energy gradients no reason for the chirality to have become specified by an "accident" that "froze up" at a hypothesized unique naissance of life. The principle of increasing entropy alone does not indicate any particular preference for the specific choice of L-amino acids and D-sugars. It may depend on some mechanism, e.g. recently it has been shown that salt-induced peptide formation reaction has a preference for the L-form of several amino acids [35]. Provided with data for thermodynamic parameters analyses using eqs. 1–4 is applicable to any mechanism to deduce energetic preferences or to simulate courses of chemical reactions [17,19]. Since we did not assume any particular mechanism we reason that the choice for L-amino acids and D-sugars in nucleic acids could have been a random choice that gradually solidified as the system grew larger. The principle of increasing entropy alone cannot argue for or against the connection between handedness of biological macromolecules and parity violation [4,35,36]. It depends on the particular mechanisms how important the minute difference between L and D forms is. In any case had two

"mirror" systems developed in isolation and confronted each other later the "battle" over matter and energy for entropy would inevitably lead to a unanimous victory of the more potent one.

The value of "standardization" is large when the base constituents are used extensively to produce functional entities. Therefore it is natural that especially the long and numerous proteins and nucleic acids are built from unique stereoisomeric residues. Systems with homochiral protein and nucleic acid polymers evolved because the ubiquitous handedness yielded more effective functions to diminish potential energy difference than racemic systems could emerge with. The value of homochirality in making functional proteins and nucleic acids has been well-understood e.g. helices and sheets form only from homochiral polymers. We emphasize that the functional value is inherent in the thermodynamic imperative; functional structures are the mechanisms to level differences in energy.

Furthermore the link between handedness and catalysis appears justified as both proteins and RNAs are catalysts, enzymes and ribozymes that assemble to diverse powerful mechanisms of energy transduction. A chain elongation by mere catalysis would nonetheless require an arsenal of precursors in numbers according to Eq. 4. Clearly these closely similar and often non-functional intermediates would contribute to the entropy production less than distinct and functional end-products from information-guided expressions of contemporary open systems known commonly as organisms.

**CONCLUSIONS**

The puzzle of homochirality has attracted attention since it was found by Pasteur. Chiral-pure outcomes have emerged from certain scenarios and understood as consequences of kinetics. Here we clarify *why* this happens by pointing out that the principle of increasing entropy, equivalent to diminishing differences in energy, underlies all kinetic courses and thus could be a cause of chirality consensus. Under influx of external energy systems evolve to high-entropy non-equilibrium states using mechanisms of energy transduction. The rate of entropy increase is the universal fitness criterion of natural selection among the diverse mechanisms that favors those that are most effective in leveling potential energy differences [17]. The ubiquitous handedness enables rapid synthesis of diverse metastable mechanisms to access free energy gradients to attain and maintain high-entropy non-equilibrium states. When the external energy is cut off, the energy gradient from the system to its exterior reverses and racemization will commence toward the equilibrium. Then the mechanisms of energy transduction have become improbable and will vanish since there are no gradients to replenish them.

The common consent that a racemic mixture has higher entropy than a chirally pure solution is certainly true at the stable equilibrium. Therefore high entropy is often associated with high disorder. However entropy is not an obscure logarithmic probability measure but probabilities describe energy densities and mutual gradients in energy. The local order and structure that associate with the mechanisms of energy transduction are well warranted when they allow the open system as a whole to access and level free energy gradients. Order and standards are needed to attain and maintain the high-entropy non-equilibrium states. We expect that the principle of increasing entropy accounts also for the universal genetic code to allow exchange of genetic material to thrust evolution toward new more probable states. The common chirality convention is often associated with a presumed unique origin of life but it reflects more the all-encompassing unity of biota on Earth that emerged from evolution over the eons.

**REFERENCES**

[1]   Cline DB. Physical Origin of Homochirality in Life. *AIP Conference Proceedings 379*. Woodbury, N.Y. (1996).

[2]   Hegstrom RA, Rein DW, Sanders PGH. Calculation of the parity nonconserving energy difference between mirror-image molecules. J Chem Phys 1980; 73: 2329–41.

[3]   Hegstrom RA. *β* Decay and the origins of biological chirality: theoretical results. Nature 1982; 297: 643–47.

[4]   Mason SF, Tranter GE. Energy inequivalence of peptide enantiomers from parity non-conservation. J Chem Soc Chem Commun 1983; 117–19.

[5]   Kondepudi DK, Nelson GW. Weak neutral currents and the origin of biomolecular chirality. Nature 1985; 314: 438–41.

[6]   Chyba C, Sagan C. Endogenous production, exogenous delivery and impact-shock synthesis of organic molecules: an inventory for the origins of life. Nature 1992; 355: 125–32.

[7] Bonner WA. The origin and amplification of biomolecular chirality. Orig Life Evol Biosph 1991; 11: 59–111.

[8] Monod J. Le hazard et la nécessité. Éditions du Seuil. Paris; Change and Necessity. New York, Knopf 1971 and London, Collins 1972.

[9] Cairns-Smith AG. Genetic Takeover and the Mineral Origins of Life. Cambridge University Press 1982.

[10] Tranter GE. Parity-violating energy differences of chiral minerals and the origin of biomolecular homochirality. Nature 1985; 318: 172–3.

[11] Frank F. On spontaneous asymmetric synthesis. Biochem Biophys Acta 1953; 11: 459–63.

[12] Shibata T, Morioka H, Hayase T, Choji K, Soai K. Asymmetric autocatalytic reaction of 3-quinolylalkanol with amplification of enantiomeric excess. J Am Chem Soc 1996; 118: 471–2.

[13] Mathew SP, Iwamura H, Blackmond DG. Amplification of enantiomeric excess in a proline-mediated reaction. Angew Chem Int Ed Engl 2004; 43: 3318–21.

[14] Cintas P. Chirality of Living Systems: A Helping Hand from Crystals and Oligopeptides. Angew Chem Int Ed Engl 2002; 41: 1139–45.

[15] Kondepudi DK, Kaufman R, Singh N. Chiral Symmetry Breaking in Sodium Chlorate Crystallizaton. Science 1990; 250: 975–6.

[16] Viedma C. Enantiomeric crystallization from D,L-aspartic and D,L-glutamic acids: implications for biomolecular chirality in the origin of life. Orig Life Evol Biosph 2001; 31: 501–9.

[17] Sharma V, Annila A. Natural Process – Natural Selection. Biophys Chem 2007; 127: 123–8.

[18] Darwin C. On the Origin of Species. London, John Murray 1859.

[19] Grönholm T, Annila A. Natural Distribution. Math Biosci 2007; in press. doi:10.1016/j.mbs.2007.07.004

[20] Kondepudi D, Prigogine I. Modern Thermodynamics. New York, Wiley 1998.

[21] Gibbs JW. The Scientific Papers of J. Willard Gibbs. Ox Bow Press 1993–1994.

[22] Atkins P, De Paula J. Physical Chemistry, 8th edition. Oxford University Press 2006.

[23] Schrödinger E. What is Life? The Physical Aspects of the Living Cell. Cambridge University Press, Cambridge 1948.

[24] Brillouin L. Science and Information Theory. New York, Academic Press 1963.

[25] Waage P, Guldberg CM. Forhandlinger 35. Videnskabs-Selskabet i Christiana, Christiana 1864.

[26] Eigen M. Steps towards life. Oxford, Oxford University Press 1992.

[27] Eigen M, Schuster P. The Hypercycle, A Principle of Natural Self-Organization. Berlin, Springer 1979.

[28] Kauffman S. Investigations. Oxford, Oxford University Press 2000.

[29] Strogatz S. Non-linear Dynamics and Chaos, Cambridge MA, Perseus Books 1994.

[30] Brandenburg A, Multamäki T. How long can left and right handed life forms coexist? Int J Astrobio 2004; 3: 209–19.

[31] Hardin G. The Competitive Exclusion Principle. Science 1960; 131, 1292-7.

[32] Kondepudi DK, Nelson GW. Chiral-symmetry-breaking states and their sensitivity in nonequilibrium chemical systems. Physica 1984; 125A: 465–96.

[33] Miller SL. A production of amino acids under possible primitive earth conditions. Science 11953; 17: 528–9.

[34] Engel MH, Macko SA, Silfer JA. Carbon isotope composition of individual amino acids in the Murchison meteorite. Nature 1990; 348: 47–9.

[35] Fitz D, Reiner H, Plankensteiner K, Rode BM. Possible Origins of Biohomochirality. Current Chemcial Biology, 2007; 1: 41-52.

[36] Bonner WA. Parity violation and the evolution of biomolecular homochirality. Chirality 2000; 12: 114–6.